\documentclass[12pt]{article}

\usepackage[english]{babel}
\usepackage{epsfig}
\usepackage{amsmath}

\usepackage{graphicx}
\usepackage{geometry}
\usepackage[latin1]{inputenc}
\usepackage{amsmath,amssymb}

\begin{document}



{\centerline{\Large{\bf{Eigenwaves in Waveguides with Dielectric
Inclusions:}}}}

{\centerline{\Large{\bf{Spectrum}}}}
\bigskip
\bigskip
{\centerline{Y. Smirnov$^{\rm a}$ and Y. Shestopalov$^{\rm b}$}}
\bigskip
{\centerline{ $^{\rm a}$ Penza State University,
Penza, 440017 Russia (smirnov@penzadom.ru) }}

{\centerline{ $^{\rm b}$ Karlstad University, SE 65188 Karlstad,
Sweden (youri.shestopalov@kau.se) }}




\bigskip
\bigskip
\noindent {\bf{Abstract.}}\quad We consider fundamental issues of
the mathematical theory of the wave propagation in waveguides with
inclusions. Analysis is performed in terms of a boundary
eigenvalue problem for the Maxwell equations
which is reduced to an eigenvalue problem for an operator pencil.
We formulate the definition of eigenwaves and associated waves
using the system of eigenvectors and associated vectors of the
pencil
and
prove
%
%
that the spectrum of normal waves forms a nonempty set of
isolated points localized in a strip with at most finitely many
real points.
\bigskip

\noindent {\bf{Keywords:}}\quad  eigenwave; waveguide; pencil;
spectrum; dielectric; inclusion

\noindent {\bf{AMS Classification:}}\quad 45E99, 31B20, 83C50,
74S10


\bigskip
\section{Introduction}
\label{section1}

Analysis of the wave propagation in waveguides with nonhomogeneous
filling and arbitrary inclusions (perfectly conducting and
dielectric) constitutes an important class of vector
electromagnetic problems. However, many urgent tasks here
remain unsolved that have been known for empty waveguides since the late 1940s; namely:
existence of
normal waves and
their basic properties including
the discreteness and localization the spectrum of normal waves on the complex plane, completeness and basis property
in terms of both longitudinal and transversal field components, and so on.

The theory of electromagnetic wave propagation in waveguides with
homogeneous filling were elaborated in classical works of A.N.
Tikhonov and A.A. Samarskii \cite{153}--\cite{155}. The analysis
in this case is reduced to two scalar selfadjoint problems which
are studied using standard methods.
List the most important results obtained for homogeneous waveguides:
there exists a (countable) set of eigenvalues (spectrum of normal waves
of a waveguide) consisting of real isolated points with the only
accumulation point at infinity and the system of normal waves is
complete and forms a basis. For nonhomogeneous waveguides with
given cross-sectional
geometry, in particular, rectangular \cite{71, 84} and circular
\cite{121}, the results concerning existence and distribution of the normal wave spectra on the complex plane are
obtained by reducing to
explicit dispersion relations and analysis of
the corresponding complex-valued functions of one or several complex variables.

To the best of our knowledge, the existence of eigenvalues and
their distribution on the complex plane remain an open issue as
well as the completeness and basis property for the system of
normal waves in nonhomogeneously filled waveguides with arbitrary
inclusions. This fact has become a main reason for us to complete
in this paper the mathematical theory of wave propagation in
waveguides by filling these gaps.

Let us briefly summarize the {\it new fundamental} results obtained in this
study for an arbitrary waveguide with nonhomogeneous filling and
arbitrary inclusions (that belongs to the considered family):

(i) the spectrum of normal waves is nonempty and forms a countable
set of isolated points on the complex plane (cut along two
intervals on the real axis) without finite accumulation points;

(ii) the spectrum is symmetric with respect to the axes on the
complex plane, is localized in a strip, and contains not more than
a finite number of real points.

During the last two decades an
increasing interest has been reported to the study of
electromagnetic wave propagation in guiding systems with
nonhomogeneous filling. Different types of them have been created
and found various practical applications and many their physical
properties have been established. Simultaneously, the interest in
developing rigorous mathematical techniques has never vanished.
A driving force here is the necessity of designing new
guiding systems such as complicated volume and planar microstrip and slot
transmission lines where nonhomogeneous structure of the guide plays the crucial role. Note also that the
study of the wave propagation in waveguides with inhomogeneous
filling requires (and leads to) elaboration of special methods of
the spectral theory of operator-valued functions (OVFs) and
operator pencils.

The typical settings that arise in mathematical models of the wave
propagation in nonhomogeneous waveguides are nonselfadjoint
boundary eigenvalue problems for the system of Helmholtz equations
with piecewise constant coefficients.
On the medium discontinuity lines (or surfaces) the transmission
conditions are added. An important feature is that the  spectral
parameter enters both the equations and transmission conditions in
a nonlinear manner. A huge amount of publications is devoted to
investigations of these problems.
However, the main attention was paid to
numerical determination
of dominant modes propagating in waveguides of various structure;
many references can be found in \cite{81, 84, 87, 104, 121,
192}.

Analysis of the  propagation of normal waves in waveguides with
nonhomogeneous filling is reduced to a vector nonselfadjoint
boundary value problem.  Complex waves may exist in such
waveguides that correspond to eigenvalues which are neither purely
real nor purely imaginary. This phenomenon was discovered and
studied in \cite{58, 70}. The existence of eigenvalues of
multiplicity greater than 1 was discussed in \cite{93, 115}.

Important contribution to the mathematical theory of
electromagnetic wave propagation in waveguides of complicated
structure was made by A.S. Ilinski and Yu.V. Shestopalov in
\cite{102}--\cite{104} and \cite{192}--\cite{194}. They propose
the reduction of the problem on normal waves in a waveguide to a
problem on characteristic numbers for a meromorphic OVF nonlinear
with respect to the spectral parameter; in the majority of cases
OVF is an operator of a system of integral equations with
logarithmic singularity of the kernel.
This approach was developed also
by E.V. Chernokozhin \cite{101} and in \cite{96}-\cite{100}.
Using this technique, the discreteness of the spectrum of normal
waves was proved for a wide family of waveguides with
nonhomogeneous filling. For slot transmission lines
the existence of eigenvalues was established
in \cite{104} . Localization of eigenvalues on
the complex plane were studied in \cite{98, 101, 103, 104}.
Note however that it is hardly possible to prove the existence and
determine the spectrum location on the complex plane by these
methods for a wide family of nonhomogeneously filled waveguides
considered in this study.

An approach based on the reduction to eigenvalue problems for
operator pencils considered in Sobolev spaces was proposed by Yu.G. Smirnov in
\cite{165, 174, 163}. General theory of polynomial
operator-functions called operator pencils is sufficiently well
elaborated
in \cite{54, 76, 77, 78, 113, 122} and \cite{125}-\cite{128}. A
fundamental work by Keldysh \cite{108} pioneered investigation of
nonselfadjoint polynomial pencils.  Note that the theory of
operator pencils is very close to the theory of nonselfadjoint
operators \cite{54, 76} and allows one to apply powerful methods
of the latter. Operator pencils were applied to the analysis of
electromagnetic problems in \cite{82, 89, 114}.

We see that the method of operator pencils
has proved to be a natural and efficient approach for
investigation of the wave propagation in waveguides. The reduction
of boundary eigenvalue problems to eigenvalue problems for
operator pencils allows one to apply various well-developed
methods of functional analysis \cite{146} in order to study
spectral properties of the pencil. This method is applied in the
present study.

Let us give a brief insight into the contents of this work.
In
Section \ref{section2} we describe a class of waveguides under consideration
and formulate the problem on normal waves for homogeneous Maxwell
equations stated in terms of longitudinal components of
electromagnetic field.
We perform the reduction to a boundary
eigenvalue problem for the system of Helmholtz equations and
introduce the notion of (weak) solution
using variational relations in Sobolev spaces.  Among characteristic features of the problem
note that the spectral parameter enters the transmission
conditions in nonlinear manner, waveguides are filled with
nonhomogeneous media, and the boundary has `edges'.  Therefore, a
special definition of the solution is required. We formulate this
definition using variational relations.

In Sections \ref{subsec:mylabel2w}  and \ref{subsec:mylabel3w} the problem is reduced to the
study of an operator pencil $L\left( \gamma \right)$ of the fourth order. We investigate
the properties of the operators of the pencil and establish basic properties of its spectrum
showing among
all that the pencil does not belong to the families of Keldysh
pencils or hyperbolic pencils.
Finally
we prove fundamental theorems concerning the
discreteness and localization of the spectrum
on the complex plane.



%

The techniques used in this study  are mainly based on the
approaches and results employing the methods of nonselfadjoint
OVFs and operator pencils published in \cite{162, 163, 174}.

\section{Statement of the problem on normal waves in a waveguide}
\label{section2}

Let $Q \subset R^2 = \left\{ {x_3 = 0} \right\}$ be a bounded
domain on the plane $Ox_1 x_2 $ with boundary $\partial Q$. Let
${\rm l} \subset Q$ be a simple closed or unclosed $C^\infty
$-smooth curve without points of intersection, dividing $Q$ into
domains $\Omega _1 $ and $\Omega _2 $; $Q = \Omega _1 \cup \Omega
_2 \cup {\rm l}$. If ${\rm l}$ is an unclosed curve, then the
points $\partial {\rm l}$ do not coincide and belong to $\partial
Q$: $\partial {\rm l} \subset \partial Q$. We will assume also
that boundaries $\partial Q$, $\partial \Omega _1 $, and $\partial
\Omega _2 $ of domains $Q$, $\Omega _1 $, and $\Omega _2 $ are
simple closed piecewise smooth curves formed by a finite number of
$C^\infty $-smooth arcs intersecting at nonvanishing angles.



\begin{center}
\begin{figure}[htbp]
\centerline{\includegraphics[width=2.28in,height=1.19in]{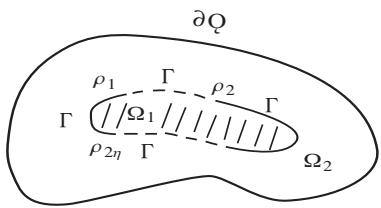}}
\caption{  \label{fig1}  Geometry of the waveguide cross section
of the first type. $\Gamma _0 = {\Gamma }' \cup \partial Q$.
$\Gamma \cup {\Gamma }'\mathop \cup \limits_i \rho_i = {\rm l}$
$\left({{\Gamma }' \cap \Gamma = \emptyset } \right)$, $\Omega
=\Omega_1 \cup \Omega _2 \cup \Gamma $.  }
\end{figure}
\end{center}

Let $\rho_i \in {\rm l}$ be arbitrary $2N$ points $\rho_i \ne
\rho_j $ dividing ${\rm l}$ into parts $\Gamma $ and $\Gamma '$
such that $\Gamma = {\rm l}\backslash \overline {\Gamma '} $,
$\Gamma ' = {\rm l}\backslash \overline \Gamma $, $\Gamma \cup
\Gamma '\mathop \cup \limits_i \rho_i = {\rm l}$ ($\Gamma \cap
\Gamma ' = \emptyset )$. If $N = 0$ then $\Gamma = {\rm l}$,
$\Gamma ' = \emptyset $. We will also use the notation $\Omega =
\Omega _1 \cup \Omega _2 \cup \Gamma $ and $\Gamma_0 =\partial Q
\cup \Gamma '$.

In the general case boundary $\partial \Omega $ of domain $\Omega
$ contains the points with inner angles $0 < \alpha \le 2\pi $. If
$\alpha = 2\pi $, such a point is called edge. Domain $\Omega $
satisfies the cone property
which allows us to apply the
embedding and trace theorems in Sobolev spaces \cite{1, 124}.

\begin{center}
\begin{figure}[htbp]
\centerline{\includegraphics[width=2.83in,height=1.39in]{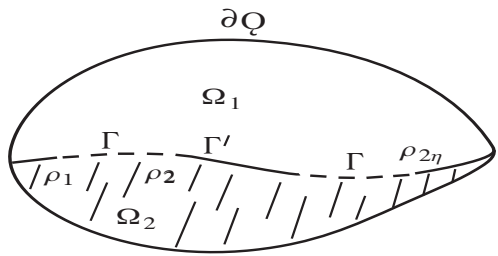}}
\caption{  \label{fig2}  Geometry of the waveguide cross section
of the second type.
}
\end{figure} 
\end{center}

We will consider the problem on normal waves in a cylindrical
shielded waveguide whose transversal (with respect to $Ox_3 )$
cross-section is formed by domain $Q$. We will assume also that
waveguide's filling contains two isotropic media with relative
permittivity $\varepsilon _j $ in domain $\Omega _j $;
$\varepsilon _j \ge 1$, $Im\,\varepsilon _j = 0$, and $\mu_j = 1$
($j = 1,2)$. Here $\Gamma _0 $ is the projection of the surface of
the infinitely thin and perfectly conducting shields and $\Gamma $
is the projection of the dielectric surfaces.

This family of waveguides contains in particular all types of
shielded transmission lines: cylindrical and rectangular
waveguides with partial filling, slot lines and  strip lines with
several slots or strips placed on a curved interface etc.
\cite{87}.




Propagation of electromagnetic waves in a guiding system is
described by the homogeneous system of Maxwell equations with
dependence $e^{i\gamma x_3 }$ on longitudinal coordinate $x_3 $
\cite{104}:
\begin{equation}
\label{eq1w}
\begin{array}{l}
 rot\,{\rm {\bf E}} = - i{\rm {\bf H}},\;X = \left( {x_1 ,\;x_2 ,\;x_3 }
\right) \in \Sigma , \\
 rot\,{\rm {\bf H}} = i\varepsilon {\rm {\bf E}},\;x = \left( {x_1 ,\;x_2 }
\right), \\
 {\rm {\bf E}}\left( X \right) = \left( {E_1 \left( x \right){\rm {\bf e}}_1
+ E_2 \left( x \right){\rm {\bf e}}_2 + E_3 \left( x \right){\rm {\bf e}}_3
} \right)e^{i\gamma x_3 }, \\
 {\rm {\bf H}}\left( X \right) = \left( {H_1 \left( x \right){\rm {\bf e}}_1
+ H_2 \left( x \right){\rm {\bf e}}_2 + H_3 \left( x \right){\rm {\bf e}}_3
} \right)e^{i\gamma x_3 }, \\
 \end{array}
\end{equation}
with the boundary conditions for the tangential electric field
components on the perfectly conducting surfaces
\begin{equation}
\label{eq2w} \left. {{\rm {\bf E}}_t } \right|_M = 0,
\end{equation}
the transmission conditions for the tangential electric and
magnetic field components on the interface (surfaces where the
permittivity is discontinuous)
\begin{equation}
\label{eq3w}
 \left[ {{\rm {\bf E}}_t } \right]_L = 0, \qquad
 \left[ {{\rm {\bf H}}_t } \right]_L = 0,
\end{equation}
and the condition that provides finiteness of energy
\begin{equation}
\label{eq4w} \int\limits_V {\left( {\varepsilon \left| {\rm {\bf
E}} \right|^2 + \left| {\rm {\bf H}} \right|^2} \right)dX} <
\infty .
\end{equation}

Here $M = \left\{ {X:x \in \Gamma _0 } \right\}$ is the shielded
part of the boundary, $L = \left\{ {X:\,x \in \Gamma } \right\}$
is the boundary where the permittivity undergoes breaks, and $V
\subset \Sigma = \left\{ {X:x \in \Omega } \right\}$ is an
arbitrary bounded domain. System of Maxwell equations (\ref{eq1w})
is written in the normalized form and we use the following
dimensionless variables and parameters \cite{63, 100}: $k_0 x \to
x$, $\sqrt {\mu _0 / \varepsilon _0 } {\rm {\bf H}} \to {\rm {\bf
H}}$, ${\rm {\bf E}} \to {\rm {\bf E}}$; $k_0^2 = \varepsilon _0
\mu _0 \omega ^2$, where $\varepsilon _0 $ and $\mu _0 $ are
permittivity  and permeability of vacuum (the time factor
$e^{i\omega t}$ is omitted).


The problem on normal waves is an eigenvalue problem for the
system of Maxwell equations with respect to spectral parameter
$\gamma $. Eigenfunctions corresponding to certain complex values
of a longitudinal wave number $\gamma $ are usually called the
normal waves of the waveguide.

Write system of Maxwell equations (\ref{eq1w}) in the form
\[
\frac{\partial H_3 }{\partial x_2 } - i\gamma H_2 = i\varepsilon E_1 ,
\quad
\frac{\partial E_3 }{\partial x_2 } - i\gamma E_2 = - iH_1 ,
\quad
i\gamma H_1 - \frac{\partial H_3 }{\partial x_1 } = i\varepsilon E_2 ,
\]
\[
i\gamma E_1 - \frac{\partial E_3 }{\partial x_1 } = - iH_2 ,
\quad
\frac{\partial H_2 }{\partial x_1 } - \frac{\partial H_1 }{\partial x_2 } =
i\varepsilon E_3 ,
\quad
\frac{\partial E_2 }{\partial x_1 } - \frac{\partial E_1 }{\partial x_2 } =
- iH_3 ,
\]
and express functions $E_1 $, $H_1 $, $E_2 $, and $H_2 $ via $E_3
$ and $H_3 $ from the first, second, fourth, and fifth equalities
\begin{equation}
\label{eq5w} E_1 = \frac{i}{\tilde {k}^2}\left( {\gamma
\frac{\partial E_3 }{\partial x_1 } - \frac{\partial H_3
}{\partial x_2 }} \right), \quad E_2 = \frac{i}{\tilde
{k}^2}\left( {\gamma \frac{\partial E_3 }{\partial x_2 } +
\frac{\partial H_3 }{\partial x_1 }} \right),
\end{equation}
\[
H_1 = \frac{i}{\tilde {k}^2}\left( {\varepsilon \frac{\partial E_3
}{\partial x_2 } + \gamma \frac{\partial H_3 }{\partial x_1 }} \right),
\quad
H_2 = \frac{i}{\tilde {k}^2}\left( { - \varepsilon \frac{\partial E_3
}{\partial x_1 } + \gamma \frac{\partial H_3 }{\partial x_2 }} \right);
\tilde {k}^2 = \varepsilon - \gamma ^2.
\]
Note that this representation is possible if $\gamma ^2 \ne
\varepsilon _1 $ and $\gamma ^2 \ne \varepsilon _2 $.

It follows from (\ref{eq5w}) that the field of a normal wave can
be expressed via two scalar functions
\[
\Pi \left( {x_1 ,x_2 } \right) = E_3 \left( {x_1 ,x_2 } \right),
\quad
\Psi \left( {x_1 ,x_2 } \right) = H_3 \left( {x_1 ,x_2 } \right).
\]

Thus the problem on normal waves is reduced to a boundary
eigenvalue problem for functions $\Pi $ and $\Psi $. Let us write
down this problem.
%
From (\ref{eq1w}) and (\ref{eq2w})  we have the following
eigenvalue problem: to find $\gamma \in C$ called eigenvalues
such that there exist nontrivial solutions of the Helmholtz
equations
\[
\Delta \Pi + \tilde {k}^2\Pi = 0,
\quad
x = \left( {x_1 ,x_2 } \right) \in \Omega _1 \cup \Omega _2 ,
\]
\begin{equation}
\label{eq6w} \Delta \Psi + \tilde {k}^2\Psi = 0, \quad \tilde
{k}^2 = \tilde {k}_j^2 = \varepsilon _j - \gamma ^2,
\end{equation}
satisfying the boundary conditions on $\Gamma _0 $
\begin{equation}
\label{eq7w} \left. \Pi \right|_{\Gamma _0 } = 0,\quad \left.
{\frac{\partial \Psi }{\partial n}} \right|_{\Gamma _0 } = 0,
\end{equation}
the transmission conditions on $\Gamma $
\begin{equation}
\label{eq8w}
\left[ \Pi \right]_\Gamma = 0,
\,\,
\left[ \Psi \right]_\Gamma = 0,
\quad
\gamma \left[ {\frac{1}{\tilde {k}^2}\frac{\partial
\Psi }{\partial \tau }} \right]_\Gamma + \left[ {\frac{\varepsilon
}{\tilde {k}^2}\frac{\partial \Pi }{\partial n}} \right]_\Gamma =
0,\,\,
\gamma \left[ {\frac{1}{\tilde {k}^2}\frac{\partial \Pi
}{\partial \tau }} \right]_\Gamma - \left[ {\frac{1}{\tilde
{k}^2}\frac{\partial \Psi }{\partial n}} \right]_\Gamma = 0,
\end{equation}
and the energy (`edge') condition
\begin{equation}
\label{eq9w} \int\limits_\Omega {(\left| {\nabla \Pi } \right|^2 +
\left| {\nabla \Psi } \right|^2 + \left| \Pi \right|^2 + \left|
\Psi \right|^2)dx} < \infty .
\end{equation}
Here $n$ and $\tau $ denote the (exterior to $\Omega_2$) normal
and tangential unit vectors such that $x_1 \times x_2 = \tau
\times n$. Square brackets $\left[ f \right]_\Gamma = \left. {f_2
} \right|_\Gamma - \left. {f_1 } \right|_\Gamma $ denote the
difference of limiting values of a function on $\Gamma $ in
domains $\Omega_2 $ and $\Omega_1 $. Conditions (\ref{eq7w}) are
to be satisfied on both sides of the part of boundary $\Gamma'$.

In order to obtain (\ref{eq6w})--(\ref{eq9w}) we used formulas
(\ref{eq5w}). Conditions (\ref{eq7w})--(\ref{eq9w}) are another
form of conditions (\ref{eq2w})--(\ref{eq4w}). Thus the
longitudinal components of a normal wave satisfy
(\ref{eq6w})--(\ref{eq9w}). The inverse assertion is true. If $\Pi
$ and $\Psi $ is a solution of problem (\ref{eq6w})--(\ref{eq9w})
then the transversal components can be determined by (\ref{eq5w}).
The field $E$, $H$ will satisfy all conditions (\ref{eq1w}) and
(\ref{eq2w})--(\ref{eq4w}). The equivalence of the reduction to
problem (\ref{eq6w})--(\ref{eq9w}) is not valid only for $\gamma
^2 = \varepsilon _1 $ or $\gamma ^2 = \varepsilon _2 $; in this
case it is necessary to study system (\ref{eq1w}) directly.

System of equations (\ref{eq6w}) with boundary conditions
(\ref{eq7w}), transmission conditions (\ref{eq8w}), and condition
(\ref{eq9w}) constitutes a boundary eigenvalue problem that will
be a subject of our study.  Note that coefficient $\varepsilon $
is not continuous and the transmission conditions contain spectral
parameter $\gamma $. Moreover, boundary $\Gamma $ may have
`edges'.

Let us formulate a definition of the solution to problem
(\ref{eq6w})--(\ref{eq9w}) that will be used in the further
analysis.

We will look for solutions to problem (\ref{eq6w})--(\ref{eq9w})
in the Sobolev spaces \cite{1, 124}
\[
H_0^1 \left( \Omega \right) = \left\{ {f:f \in H^1\left( \Omega
\right),\,\,\,\left. f \right|_{\Gamma _0 } = 0} \right\}, \quad
\mathord{\buildrel{\lower3pt\hbox{$\scriptscriptstyle\frown$}}\over
{H}} ^1\left( \Omega \right) = \left\{ {f:f \in H^1\left( \Omega
\right),\,\,\,\int\limits_\Omega {fdx} = 0} \right\}
\]
with the inner product and the norm
\[
\left( {f,g} \right)_1 = \int\limits_\Omega {\nabla f\nabla \bar {g}dx} ,
\quad
\left\| f \right\|_1^2 = \left( {f,f} \right)_1 .
\]

The seminorm $\left\| {\, \cdot \,} \right\|_1 $ in $H^1\left(
\Omega \right)$ is a norm in $H_0^1 \left( \Omega \right)$ and
$\mathord{\buildrel{\lower3pt\hbox{$\scriptscriptstyle\frown$}}\over
{H}} ^1\left( \Omega \right)$ because sesquilinear form $\left(
{f,g} \right)_1 $ in these spaces is coercive \cite{1}. Note that
it is sufficient to use the boundedness of $\Omega $ in order to
prove that the form is coercive in $H_0^1 \left( \Omega \right)$;
however it is necessary to use the cone property for the proof of
the coercive property of the form in
$\mathord{\buildrel{\lower3pt\hbox{$\scriptscriptstyle\frown$}}\over
{H}} ^1\left( \Omega \right))$. Spaces $H_0^1 \left( \Omega
\right)$ and
$\mathord{\buildrel{\lower3pt\hbox{$\scriptscriptstyle\frown$}}\over
{H}} ^1\left( \Omega \right)$ can be defined as a supplement of
spaces of infinitely smooth functions $C_0^\infty \left( \Omega
\right)$ and $C^\infty \left( \Omega \right)$ with respect to the
norm $\left\| {\, \cdot \,} \right\|_1 $ (under the condition
$\left\| f \right\|_1 < \infty )$;
$\mathord{\buildrel{\lower3pt\hbox{$\scriptscriptstyle\frown$}}\over
{H}} ^1\left( \Omega \right)$ is a subspace of functions from
$H^1\left( \Omega \right)$ which are orthogonal to the unit
function.

Under the above assumptions the domain $\Omega $ satisfies cone property:
there is a cone
\[
K_0 = \left\{ {x:0 \le x_1 \le b,\,\,x_2^2 \le ax_1^2 ;\,\,a > 0,\,\,b > 0}
\right\}
\]
such that any point $P \in \Omega $ can be a vertex of cone $K_p$
which is equal to $K_0$, and the cone belongs to $\Omega $, $K_p
\subset \Omega $. This property allows us to apply the Sobolev
trace theorem \cite{1} and consider the trace of function $f \in
H^1\left( \Omega \right)$ on $\Gamma $ as an element of space
$H^{1 / 2}\left( \Gamma \right)$. Due to the trace theorem, the
relation $\left. f \right|_{\Gamma _0 } = 0$ means that the
function is equal to zero in $H^{1 / 2}\left( {\Gamma _0 }
\right)$. For any function $f \in H^1\left( \Omega \right)$ we
have $\left[ f \right]_\Gamma = 0$ in the sense of space $H^{1 /
2}\left( \Gamma \right)$; and {\it {vice versa}}, if $\left[ f
\right]_\Gamma = 0$, $\left. f \right|_{\Omega _1 } \in H^1\left(
{\Omega _1 } \right)$, $\left. f \right|_{\Omega _2 } \in
H^1\left( {\Omega _2 } \right)$, then $f \in H^1\left( \Omega
\right)$. On the part of the boundary $\Gamma ' \subset \Gamma _0
$ the trace theorem should be applied on both sides of $\Gamma '$;
in this case functions $f \in H^1\left( \Omega \right)$ have in
general different traces on different sides of $\Gamma '$. Note
also that the following embeddings
\[
H_0^1 \left( \Omega \right) \subset H_0^1 \left( Q \right) \subset H^1\left(
Q \right) \subset H^1\left( \Omega \right),
\]
hold but all embeddings are not dense if $\Gamma ' \ne \emptyset
$.

Assume that $\Pi \in H_0^1 \left( \Omega \right)$, $\Psi \in
\mathord{\buildrel{\lower3pt\hbox{$\scriptscriptstyle\frown$}}\over
{H}} ^1\left( \Omega \right)$. Condition (\ref{eq6w}) is fulfilled
in $\Omega _1 $ and $\Omega _2 $ in terms of distributions
\cite{123}. Moreover, we have for the boundary condition on
$\Gamma _0$
\[
\left. {\Pi _j } \right|_{\Gamma _0 \cap \partial \Omega _j } \in H^{1 /
2}\left( {\Gamma _0 \cap \partial \Omega _j } \right),
\quad
\left. {\frac{\partial \Psi _j }{\partial n}} \right|_{\Gamma _0 \cap
\partial \Omega _j } \in H^{ - 1 / 2}\left( {\Gamma _0 \cap \partial \Omega
_j } \right).
\]
For the transmission condition on $\Gamma $, we have
\[
\left. \Pi \right|_\Gamma \in H^{1 / 2}\left( \Gamma \right),
\quad
\left. \Psi \right|_\Gamma \in H^{1 / 2}\left( \Gamma \right),
\]
\[
\left. {\frac{\partial \Pi _j }{\partial n}} \right|_\Gamma ,\,\,\left.
{\frac{\partial \Psi _j }{\partial n}} \right|_\Gamma \in H^{ - 1 / 2}\left(
\Gamma \right),
\quad
\left. {\frac{\partial \Pi }{\partial \tau }} \right|_\Gamma ,\,\,\left.
{\frac{\partial \Psi }{\partial \tau }} \right|_\Gamma \in H^{ - 1 /
2}\left( \Gamma \right),
\]
where $\Pi_j $ and $\Psi_j$ are restrictions of $\Pi $ and $\Psi $
on $\Omega _j$.

Let us give a variational formulation of problem
(\ref{eq3w})--(\ref{eq9w}). Multiply equations (\ref{eq3w}) and
(\ref{eq4w}) by arbitrary test functions $\bar {u} \in H_0^1
\left( \Omega \right)$ and $\bar {v} \in
\mathord{\buildrel{\lower3pt\hbox{$\scriptscriptstyle\frown$}}\over
{H}} ^1\left( \Omega \right)$ (we may assume that these functions
are continuously differentiable in $\bar {\Omega }_1 $ and $\bar
{\Omega }_2 $ because these spaces form dense sets in $H_0^1
\left( \Omega \right)$ and
$\mathord{\buildrel{\lower3pt\hbox{$\scriptscriptstyle\frown$}}\over
{H}} ^1\left( \Omega \right))$, and apply Green's formula \cite{123},
\cite{6}
for each domain $\Omega _j $ separately. Note that the possibility
of applying Green's formula for these functions is proved in \cite{123} and
\cite{6}, p.~618. We have
\begin{equation}
\label{eq10w} \int\limits_{\Omega _j } {\nabla \Pi \nabla \bar
{u}dx} - \tilde {k}_j^2 \int\limits_{\Omega _j } {\Pi \bar {u}dx}
= \left( { - 1} \right)^j\int\limits_{\partial \Omega _j } {\left.
{\frac{\partial \Pi }{\partial n}} \right|_{\partial \Omega _j }
\bar {u}d\tau } ,
\end{equation}
\begin{equation}
\label{eq11w} \int\limits_{\Omega _j } {\nabla \Psi \nabla \bar
{v}dx} - \tilde {k}_j^2 \int\limits_{\Omega _j } {\Psi \bar {v}dx}
= \left( { - 1} \right)^j\int\limits_{\partial \Omega _j } {\left.
{\frac{\partial \Psi }{\partial n}} \right|_{\partial \Omega _j }
\bar {v}d\tau } ;\,\,\,j = 1,2.
\end{equation}
Then, substituting the normal derivatives from (\ref{eq7w}) and
(\ref{eq8w}) to (\ref{eq10w}) and (\ref{eq11w}) we obtain the
variational relation
\begin{equation}
\label{eq12w}
 \int\limits_\Omega {\frac{1}{\tilde {k}^2}\left( {\varepsilon \nabla \Pi
\nabla \bar {u} + \nabla \Psi \nabla \bar {v}} \right)dx} -
\int\limits_\Omega {\left( {\varepsilon \Pi \bar {u} + \Psi \bar {v}}
\right)dx}
- \gamma \left[ {\frac{1}{\tilde {k}^2}}
\right]\int\limits_\Gamma {\left( {\frac{\partial \Pi }{\partial \tau }\bar
{v} - \frac{\partial \Psi }{\partial \tau }\bar {u}} \right)d\tau } = 0, \\
\end{equation}
which is derived for smooth functions $u$, $v$. In Section 2.1 we
will prove the continuity of the sesquilinear forms defined by the
integrals in (\ref{eq12w}). Hence relation (\ref{eq12w}) can be
extended to arbitrary functions $u \in H_0^1 \left( \Omega
\right)$, $v \in
\mathord{\buildrel{\lower3pt\hbox{$\scriptscriptstyle\frown$}}\over
{H}} ^1\left( \Omega \right)$. Here and below, the $f$ under the
integral sign in $\int\limits_\Gamma {fd\tau } $ is the trace of
the function on $\Gamma $.

For $v \equiv 1$, $u \equiv 0$ we obtain in a similar manner
\begin{equation}
\label{eq13w} \int\limits_\Omega {\Psi dx} = - \int\limits_\Gamma
{\left[ {\frac{1}{\tilde {k}^2}\frac{\partial \Psi }{\partial n}}
\right]_\Gamma d\tau } = - \gamma \left[ {\frac{1}{\tilde {k}^2}}
\right]\int\limits_\Gamma {\left. {\frac{\partial \Pi }{\partial
\tau }} \right|_\Gamma d\tau } = 0;
\end{equation}
consequently, the choice of space
$\mathord{\buildrel{\lower3pt\hbox{$\scriptscriptstyle\frown$}}\over
{H}} ^1\left( \Omega \right)$ does not contradict to the choice of
the space of solutions to problem (\ref{eq6w})--(\ref{eq9w}). In
(\ref{eq13w}) we used the condition
\[
\left. \Pi \right|_{\rm l} \in \tilde {H}^{1 / 2}\left( \bar {\Gamma } \right):= \{\varphi : \varphi \in
H^{1/2}({\rm l}), \quad  \mbox{supp} \quad \varphi \subset \bar {\Gamma }\},
\]
since $\Pi \in H_0^1 \left( \Omega \right) $, so that the set of
functions $C_0^\infty \left( \Gamma \right)$ is dense in $\tilde
{H}^{1 / 2}\left( \bar {\Gamma } \right)$.

\textbf{Definition 1}. \textit{
\label{def1} The pair of functions}

$\Pi \in H_0^1 \left( \Omega \right)$,
$\Psi \in
\mathord{\buildrel{\lower3pt\hbox{$\scriptscriptstyle\frown$}}\over
{H}}^1\left( \Omega \right) $   $(\left\| \Pi \right\|_1 + \left\|
\Psi \right\|_1 \ne 0)$

\noindent \textit{ is called the eigenvector of problem
(\ref{eq6w})--(\ref{eq9w}) corresponding to eigenvalue $\gamma_0$
if variational relation (\ref{eq12w}) holds for} $u\in
H_0^1(\Omega)$, $ v
\in\mathord{\buildrel{\lower3pt\hbox{$\scriptscriptstyle\frown$}}\over
{H}}^1\left( \Omega \right)$.

Thus, if $\Pi \in H_0^1 \left( \Omega \right)$ and $\Psi \in
\mathord{\buildrel{\lower3pt\hbox{$\scriptscriptstyle\frown$}}\over
{H}} ^1\left( \Omega \right)$ and (\ref{eq6w})--(\ref{eq9w}) are
fulfilled, variational relation (\ref{eq12w}) also holds. The
inverse assertion is true. Choosing $u$ and $v$ with a support in
$\Omega _j $ we have that equations (\ref{eq6w}) are fulfilled in
terms of distributions. The first condition in (\ref{eq7w}), the
first condition in (\ref{eq8w}), and (\ref{eq9w}) are fulfilled by
the definition of spaces $H_0^1 \left( \Omega \right)$ and
$\mathord{\buildrel{\lower3pt\hbox{$\scriptscriptstyle\frown$}}\over
{H}} ^1\left( \Omega \right)$. If we choose $u \equiv 0$ and
assume that the support of $v$ contains the part $\Gamma _1 $ of
boundary $\Gamma _0 $, then from (\ref{eq12w}) and Green's formula
we find \cite{6}, \cite{123} that the second condition in (\ref{eq7w}) is also
fulfilled in term of distributions. Choosing arbitrary $u$ and $v$
on $\Gamma $ in (\ref{eq12w}) and applying formulas (\ref{eq10w})
and (\ref{eq11w}) we obtain the relation
\[
\int\limits_\Gamma {\left( {\gamma \left[ {\frac{1}{\tilde
{k}^2}\frac{\partial \Psi }{\partial \tau }} \right]_\Gamma + \left[
{\frac{\varepsilon }{\tilde {k}^2}\frac{\partial \Pi }{\partial n}}
\right]_\Gamma } \right)\bar {u}d\tau } + \int\limits_\Gamma {\left( {\gamma
\left[ {\frac{1}{\tilde {k}^2}\frac{\partial \Pi }{\partial \tau }}
\right]_\Gamma - \left[ {\frac{1}{\tilde {k}^2}\frac{\partial \Psi
}{\partial n}} \right]_\Gamma } \right)\bar {v}d\tau = 0} ,
\]
which yields that the second and third condition in (\ref{eq8w})
are fulfilled in terms of distributions.

Let us give some remarks concerning the smoothness of eigenvectors
of problem (\ref{eq6w})--(\ref{eq9w}). It is well known \cite{120,
137} that solutions $\Pi $ and $\Psi $ of homogeneous Helmholtz
equations (\ref{eq6w}) are infinitely smooth in $\Omega _1 $ and
$\Omega _2 $: $\Pi ,\Psi \in C^\infty \left( {\Omega _1 \cup
\Omega _2 } \right)$; consequently, equations (\ref{eq6w}) are
satisfied in the classical sense. In a vicinity of an arbitrary
smooth part $\Gamma _1 $ of boundary $\Gamma _0 $ conditions
(\ref{eq7w}) are also fulfilled in the classical sense and
functions $\Pi $ and $\Psi $ are infinitely smooth up to the
boundary. The behavior of $\Pi $ and $\Psi $ close to edge (angle)
points was analyzed in \cite{111}. Note that in what follows we
will not use the smoothness of functions $\Pi $ and $\Psi $.

\section{Eigenvalue problem for operator pencil}
\label{subsec:mylabel2w}

Multiplying (\ref{eq12w}) by $\tilde {k}_1^2 \tilde {k}_2^2 $ we
rewrite it in the form
\begin{equation}
\label{eq14w}
\begin{array}{l}
 {\begin{array}{*{20}c}
 \hfill & \hfill \\
\end{array} }\gamma ^4\int\limits_\Omega {\left( {\varepsilon \Pi \bar {u} +
\Psi \bar {v}} \right)dx} + \gamma ^2\left( {\int\limits_\Omega {\left(
{\varepsilon \nabla \Pi \nabla \bar {u} + \nabla \Psi \nabla \bar {v}}
\right)dx} - } \right. \\
 \left. { - \left( {\varepsilon _1 + \varepsilon _2 }
\right)\int\limits_\Omega {\left( {\varepsilon \Pi \bar {u} + \Psi \bar {v}}
\right)dx} } \right) + \left( {\varepsilon _1 - \varepsilon _2 }
\right)\gamma \int\limits_\Gamma {\left( {\frac{\partial \Pi }{\partial \tau
}\bar {v} - \frac{\partial \Psi }{\partial \tau }\bar {u}} \right)d\tau + }
\\
 \;\; + \varepsilon _1 \varepsilon _2 \left( {\int\limits_\Omega {\left(
{\varepsilon \Pi \bar {u} + \Psi \bar {v}} \right)dx} - \int\limits_\Omega
{\left( {\nabla \Pi \nabla \bar {u} + \frac{1}{\varepsilon }\nabla \Psi
\nabla \bar {v}} \right)dx} } \right) = 0, \\
 {\begin{array}{*{20}c}
 {{\begin{array}{*{20}c}
 {\;\;\;} \hfill & \hfill & \hfill \\
\end{array} }} \hfill & \hfill & \hfill \\
\end{array} }\forall u \in H_0^1 \left( \Omega \right),\,\,v \in
\mathord{\buildrel{\lower3pt\hbox{$\scriptscriptstyle\frown$}}\over {H}}
^1\left( \Omega \right). \\
 \end{array}
\end{equation}

Let $H = H_0^1 \left( \Omega \right)\times
\mathord{\buildrel{\lower3pt\hbox{$\scriptscriptstyle\frown$}}\over {H}}
^1\left( \Omega \right)$ be the Cartesian product of the Hilbert spaces with
the inner product and the norm
\[
\left( {f,g} \right) = \left( {f_1 ,g_1 } \right)_1 + \left( {f_2 ,g_2 }
\right)_1 ,
\quad
\left\| f \right\|^2 = \left\| {f_1 } \right\|_1^2 + \left\| {f_2 }
\right\|_1^2 ,
\]
where
\[
f,g \in H,
\quad
f = \left( {f_1 ,f_2 } \right)^{\rm T},
\quad
g = \left( {g_1 ,g_2 } \right)^{\rm T},
\quad
f_1 ,g_1 \in H_0^1 \left( \Omega \right),
\quad
f_2 ,g_2 \in
\mathord{\buildrel{\lower3pt\hbox{$\scriptscriptstyle\frown$}}\over {H}}
^1\left( \Omega \right).
\]
Then the integrals in (\ref{eq14w}) can be considered as
sesquilinear forms on ${\bf C}$ defined in $H$ with respect to
vector-functions such that
\[
f_1 = \Pi , \quad f_2 = \Psi,
\quad
g_1 = u, \quad g_2 = v.
\]
These forms (if they are bounded) define, in accordance with the
results of \cite{106}, linear bounded operators $T:\,H \to H$
\begin{equation}
\label{eq15w} t\left( {f,g} \right) = \left( {Tf,g} \right), \quad
\forall g \in H.
\end{equation}
Linearity follows here from the linearity  of the form with
respect to the first argument and continuity from the
estimates
\[
\left\| {Tf} \right\|^2 = t\left( {f,Tf} \right) \le C\left\| f
\right\|\,\left\| {Tf} \right\|.
\]

Consider the following quadratic forms and corresponding operators
\[
a_1 \left( {f,g} \right): = \int\limits_\Omega {\left( {\varepsilon \nabla
f_1 \nabla \bar {g}_1 + \nabla f_2 \nabla \bar {g}_2 } \right)dx} = \left(
{A_1 f,g} \right),
\quad
\forall g \in H,
\]
\[
a_2 \left( {f,g} \right): = \int\limits_\Omega {\left( {\nabla f_1 \nabla
\bar {g}_1 + \frac{1}{\varepsilon }\nabla f_2 \nabla \bar {g}_2 } \right)dx}
= \left( {A_2 f,g} \right),
\quad
\forall g \in H,
\]
\[
k\left( {f,g} \right): = \int\limits_\Omega {\left( {\varepsilon f_1
\bar {g}_1 + f_2 \bar {g}_2 } \right)dx} = \left(
{Kf,g} \right),
\quad
\forall g \in H,
\]
\begin{equation}
\label{eq16w} s\left( {f,g} \right): = \int\limits_\Gamma {\left(
{\frac{\partial f_1 }{\partial \tau }\bar {g}_2 - \frac{\partial
f_2 }{\partial \tau }\bar {g}_1 } \right)d\tau } = \left( {Sf,g}
\right), \forall g \in H.
\end{equation}

It is easy to see that forms $a_1 \left( {f,g} \right)$ and $a_2
\left( {f,g} \right)$ are bounded. The same property for the form
$k\left( {f,g} \right)$ follows from Poincare's inequality
\cite{1}.

Let us prove that form $s\left( {f,g} \right)$ is also bounded.
Assume that functions $f_1 ,\,f_2 ,\,g_1 ,\,g_2 \in C^1\left(
{\bar {\Omega }_1 } \right) \cap C^1\left( {\bar {\Omega }_2 }
\right)$. Then
\[
\int\limits_\Gamma {\left( {\frac{\partial f_1 }{\partial \tau }\bar {g}_2 -
\frac{\partial f_2 }{\partial \tau }\bar {g}_1 } \right)d\tau } =
\int\limits_\Omega {\frac{\xi }{2}\left( {\frac{\partial f_1 }{\partial x_2
}\frac{\partial \bar {g}_2 }{\partial x_1 } - \frac{\partial f_1 }{\partial
x_1 }\frac{\partial \bar {g}_2 }{\partial x_2 } + \frac{\partial f_2
}{\partial x_1 }\frac{\partial \bar {g}_1 }{\partial x_2 } - \frac{\partial
f_2 }{\partial x_2 }\frac{\partial \bar {g}_1 }{\partial x_1 }} \right)dx}
,
\]
where
\[
\xi = \left\{ {\begin{array}{l}
 1,\,\,x \in \Omega _1 \\
 - 1,\,\,x \in \Omega _2 \\
 \end{array}} \right. .
\]
Using the Schwartz inequality we finally obtain
\begin{equation}
\label{eq17w} \left| {s\left( {f,g} \right)} \right| \le
\frac{1}{2}\left\| f \right\|\,\left\| g \right\|.
\end{equation}

The required property of the form may be easily obtained if to
extend the estimate given by (\ref{eq17w}) for arbitrary functions
$f,\,g \in H$ using the continuity.

{\it{Remark}.} In the expression
\[
\int\limits_\Gamma {\left( {\frac{\partial f_1 }{\partial \tau }\bar {g}_2 -
\frac{\partial f_2 }{\partial \tau }\bar {g}_1 } \right)d\tau }
\]
$g_1$, $g_2 $, $\frac{\partial f_1 }{\partial \tau }$, and
$\frac{\partial f_2 }{\partial \tau }$ mean the restriction of elements
\[
\left. {g_1 ,g_2 } \right|_{\partial \Omega _j } \in H^{1 / 2}\left(
{\partial \Omega _j } \right),
\quad
\frac{\partial f_1 }{\partial \tau },
\quad
\left. {\frac{\partial f_2 }{\partial \tau }} \right|_{\partial \Omega _j }
\in H^{ - 1 / 2}\left( {\partial \Omega _j } \right)
\]
on $\Gamma $ \cite{123}. Since on ${\rm l}$
\[
\left. {f_1 } \right|_{\Gamma '} = 0,
\quad
\left. {g_1 } \right|_{\Gamma '} = 0,
\quad
\mbox{supp}\;f_1 \subset \bar {\Gamma },
\quad
\mbox{supp}\;g_1 \subset \bar {\Gamma },
\]
the following formulas of the integration by parts hold
\begin{equation}
\label{eq18w} \int\limits_\Gamma {\frac{\partial f_1 }{\partial
\tau }\bar {g}_2 d\tau } = - \int\limits_\Gamma {\frac{\partial
\bar {g}_2 }{\partial \tau }f_1 d\tau } ,\;\;\int\limits_\Gamma
{\frac{\partial f_2 }{\partial \tau }\bar {g}_1 d\tau } = -
\int\limits_\Gamma {\frac{\partial \bar {g}_1 }{\partial \tau }f_2
d\tau } .
\end{equation}

Now the variational problem given by (\ref{eq14w}) can be written
in the operator form
\[
\left( {L\left( \gamma \right)f,g} \right) = 0,
\quad
\forall g \in H,
\]
which is equivalent to the following equation for the operator-valued pencil
\begin{equation}
\label{eq19w}
\begin{array}{l}
 {\begin{array}{*{20}c}
 {{\begin{array}{*{20}c}
 {\;\;\;\;\;} \hfill & \hfill & \hfill \\
\end{array} }} \hfill & \hfill & \hfill \\
\end{array} }L\left( \gamma \right)f = 0,\,\,L\left( \gamma \right):H \to H,
\\
 L\left( \gamma \right): = \gamma ^4K + \gamma ^2\left( {A_1 - \left(
{\varepsilon _1 + \varepsilon _2 } \right)K} \right) + \left( {\varepsilon
_1 - \varepsilon _2 } \right)\gamma S + \varepsilon _1 \varepsilon _2 \left(
{K - A_2 } \right), \\
 \end{array}
\end{equation}
where all the operators are bounded.

Equation (\ref{eq19w}) is another form of variational relation
(\ref{eq14w}). Eigenvalues and eigenvectors of the pencil coincide
with eigenvalues and eigenfunctions of problem
(\ref{eq6w})--(\ref{eq9w}) for $\gamma ^2 \ne \varepsilon _1 $,
$\gamma ^2 \ne \varepsilon _2 $ according to the definition.

Thus the problem on normal waves is reduced to an eigenvalue
problem for pencil $L\left( \gamma \right)$.

Now we consider the properties of the operators in (\ref{eq16w}).

\textbf{Lemma 1.} \textit{
\label{lem1}   The operators $A_1$ and $ A_2 $ are uniformly
positive:
\begin{equation}
\label{eq20w} I \le A_1 \le \varepsilon _{\max } I, \quad
\varepsilon _{\max }^{ - 1} I \le A_2 \le I,
\end{equation}
where $\varepsilon _{\max } = \max \left( {\varepsilon _1
,\varepsilon _2 } \right)$ and $I$ is the unit operator in $H$.
}

The proof of the lemma is reduced to the verification of simple
inequalities
\[
\left\| f \right\|^2 \le \left( {A_1 f,f} \right) \le \varepsilon
_{\max } \left\| f \right\|^2,\qquad
\varepsilon _{\max }^{ - 1} \left\| f \right\|^2 \le \left( {A_2 f,f}
\right) \le \left\| f \right\|^2.
\]

\textbf{Lemma 2}.
\textit{
\label{lem2} The operator $S$ is selfadjoint, $S = S^\ast $, and
the following inequalities hold
\begin{equation}
\label{eq21w}
 - \frac{1}{2}I \le S \le \frac{1}{2}I.
\end{equation}
}

The selfadjointness of operator $S$ follows from the equality
\[
\int\limits_\Gamma {\left( {\frac{\partial f_1 }{\partial \tau }\bar {g}_2 -
\frac{\partial f_2 }{\partial \tau }\bar {g}_1 } \right)d\tau } =
\int\limits_\Gamma {\left( {\frac{\partial \bar {g}_1 }{\partial \tau }f_2 -
\frac{\partial \bar {g}_2 }{\partial \tau }f_1 } \right)d\tau } ,
\]
which follows in its turn from the remark above and formulas
(\ref{eq18w}). Inequalities (\ref{eq21w}) follow from
(\ref{eq17w}).

%
\textbf{Lemma 3}. \textit{
\label{lem3} Operator $K$ is positive, $K > 0$, and compact. The
following estimate holds for its eigenvalues
\begin{equation}
\label{eq22w} \lambda _n \left( K \right) = O\left( {n^{ - 1}}
\right),\,\,\,n \to \infty .
\end{equation}
}

It is easy to see that $\left( {Kf,f} \right) > 0$ for $f \ne 0$
since $\int\limits_\Omega {\left( {\varepsilon \left| {f_1 }
\right|^2 + \left| {f_2 } \right|^2} \right)dx} = 0$ is fulfilled
only for $f_1 = 0$ and $f_2 = 0$ (in $H^1\left( \Omega \right))$.

The compactness of the operator $K$ follows from formula (\ref{eq22w}).

The proof of formula (\ref{eq22w}) is based on the
Courant variational principle. From the inequality
\[
\int\limits_\Omega {\left( {\varepsilon \left| {f_1 } \right|^2 + \left|
{f_2 } \right|^2} \right)dx} \le \varepsilon _{\max } \int\limits_\Omega
{\left( {\left| {f_1 } \right|^2 + \left| {f_2 } \right|^2} \right)dx}
\]
we obtain \cite{77} that
\[
\lambda _n \left( K \right) \le \varepsilon _{\max } \lambda _n
\left( {K_H } \right), \quad n \ge 1,
\]
where $\lambda _n \left( {K_H } \right)$ are eigenvalues of the
operator specified by the sesquilinear form
\begin{equation}
\label{eq23w} q\left( {f,g} \right): = \int\limits_\Omega {\left(
{f_1 \overline g _1 + f_2 \overline g _2 } \right)dx} = \left(
{K_H f,g} \right), \quad \forall g \in H.
\end{equation}
Thus it is sufficient to consider operator $K_H $. Formula (\ref{eq22w})
follows from the asymptotic behavior of $s$-numbers \cite{77} of operator $K_H $
which can be found in \cite{Tri} (Theorem 4.10.1).
The statement is proved. (The proof in detail one can find in \cite{162})

Thus all operators $A_1 $, $A_2 $, $K$, and $S$ are selfadjoint,
and $\mbox{Ker}K = \left\{ 0 \right\}$. There exist bounded
inverse operators $A_j^{ - 1} :H \to H$ and $A_j^{1 / 2}
,\,\,A_j^{ - 1 / 2} :H \to H$; these operators are uniformly
positive. Note that in this case the condition $B > 0$ leads to $B
= B^\ast $ since Hilbert space $H$ is considered on complex-valued
functions. Using these lemmas we obtain

{\sc {Corollary 1.}}
 \textit{Operator pencil }$L\left( \gamma \right)$\textit{ is
 self-adjoint:}
\begin{equation}
\label{eq24w}
L^\ast \left( \gamma \right) = L\left( \bar {\gamma
} \right).
\end{equation}

From variational relation (\ref{eq14w}) we obtain

{\sc {Corollary 2.}}
\textit{Let }$P$\textit{ be an operator such that}
\[
P\left( {f_1 ,f_2 } \right)^{\rm T} = \left( { - f_1 ,f_2 } \right)^{\rm T}.
\]
\textit{Then}
\[
A_1 = PA_1 P,
\quad
A_2 = PA_2 P,
\quad
K = PKP,
\quad
S = - PSP,
\]
 \textit{and the following representation holds}
\begin{equation}
\label{eq25w} L\left( { - \gamma } \right) = PL\left( \gamma
\right)P.
\end{equation}

The proof of this statement  follows directly from the explicit
form of variational relation (\ref{eq14w}).

Note that operator $S$ does not possess the Fredholm property
because $\dim \mbox{Ker}\,S = \infty $. Indeed, all functions $f =
\left( {f_1 ,f_2 } \right)^{\rm T}$ satisfying the additional conditions
$\left. {f_1 } \right|_\Gamma = 0$, $\left. {f_2 } \right|_\Gamma
= 0$ belong to the kernel of operator $S$.

\section{Properties of the spectrum of pencil $L\left( \gamma \right)$}
\label{subsec:mylabel3w}

We will denote by ${\cal R} \left( L \right)$ the resolvent set of
$L\left( \gamma \right)$ (consisting of all complex values of
$\gamma $ at which there exists a bounded inverse operator $L^{ -
1}\left( \gamma \right))$ and by $\sigma \left( L \right) = {\bf
C}\backslash {\cal R} \left( L \right)$ the spectrum of $L\left(
\gamma \right)$.

In what follows we will use the definitions of finite-meromorphic
OVFs and canonical system of eigenvectors and associated vectors
of an OVF formulated in
\cite{77, 78, 107}. We will consider OVFs
that have eigenvalues with finite algebraic multiplicity.


Let us study the spectrum of pencil $L\left( \gamma \right)$. It
is more convenient to consider a regularized pencil
\begin{equation}
\label{eq29w}
\begin{array}{c}
 \tilde {L}\left( \gamma \right): = A_1^{ - 1 / 2} L\left( \gamma
\right)A_1^{ - 1 / 2} = \gamma ^4\tilde {K} + \gamma ^2\left( {I - \left(
{\varepsilon _1 + \varepsilon _2 } \right)\tilde {K}} \right) + \\
 + \left( {\varepsilon _1 - \varepsilon _2 } \right)\gamma \tilde {S} +
\varepsilon _1 \varepsilon _2 \left( {\tilde {K} - \tilde {A}_2 } \right) \\
 \end{array},
\end{equation}
where $\tilde {K} = A_1^{ - 1 / 2} KA_1^{ - 1 / 2} $, $\tilde {S}
= A_1^{ - 1 / 2} SA_1^{ - 1 / 2} $, and $\tilde {A}_2 = A_1^{ - 1
/ 2} A_2 A_1^{ - 1 / 2} $.

It is easy to see that $\sigma \left( L \right) = \sigma \left(
\tilde {L} \right)$ and the following relations hold for
eigenvectors and associated vectors
\begin{equation}
\label{eq30w} \varphi _j \left( L \right) = A_1^{ - 1 / 2} \varphi
_j \left( \tilde {L} \right).
\end{equation}

Operators $\tilde {K}$, $\tilde {S}$ and $\tilde {A}_2 $ keep all
properties of operators $K$, $S$, and $A_2 $ given in Lemmas 1--3
with the estimates
\begin{equation}
\label{eq31w}
 - \frac{1}{2}I \le \tilde {S} \le \frac{1}{2}I,
\quad
\varepsilon _{\max }^{ - 2} I \le \tilde {A}_2 \le I.
\end{equation}

Properties of the spectrum of pencil $L\left( \gamma \right)$ are
summarized in the following theorems.

\textbf{Theorem 1}.

\label{th1}
\[
\sigma \left( L \right) \subset \Pi _l = \left\{ {\gamma :\left| {Re\gamma }
\right| < l} \right\}
\]
\textit{ i.e. for a certain $l > 0$ the spectrum of pencil
$L\left( \gamma \right)$  lies in the strip $\Pi_l$.}

{\bf{Proof.}} In order to prove this theorem, assume that $l >
\sqrt {\varepsilon _1 + \varepsilon _2 } $ and consider the
operator-valued function
\begin{equation}
\label{eq32w} F\left( \gamma \right): = \left( {\gamma ^2 - \left(
{\varepsilon _1 + \varepsilon _2 } \right)} \right)^{ - 1}\tilde
{L}\left( \gamma \right) = \gamma ^2\tilde {K} + I + \gamma ^{ -
1}T\left( \gamma \right),
\end{equation}
in the domain $D_0 = \left\{ {\gamma :\left| \gamma \right| > l} \right\}$,
where
\[
T\left( \gamma \right) = \gamma \left( {\gamma ^2 - \left( {\varepsilon _1 +
\varepsilon _2 } \right)} \right)^{ - 1}\times \left( {\left( {\varepsilon
_1 + \varepsilon _2 } \right)I + \left( {\varepsilon _1 - \varepsilon _2 }
\right)\gamma \tilde {S} + \varepsilon _1 \varepsilon _2 \left( {\tilde {K}
- \tilde {A}_2 } \right)} \right)
\]
is a holomorphic and bounded operator-valued function in the
domain $D_0 $: $\left\| {T\left( \gamma \right)} \right\| \le T_0
$ for $\gamma \in D_0 $. One can see that $\sigma \left( \tilde
{L} \right) \cap D_0 = \sigma \left( F \right) \cap D_0 $. If
$\left| {Re\gamma } \right| > l$, there exists a bounded operator
\[
R\left( \gamma \right) = \left( {\gamma ^2\tilde {K} + I} \right)^{ - 1}
\]
and its norm can be calculated by the formula (see \cite{76}, p. 309)
$$
\|R(\gamma)\| = \frac{\gamma^{-2}}{d(-\gamma^{-2})},$$
where $d(\mu)$ is the distance from point $\mu$ to spectrum of operator $\tilde K$.
Thus we obtain
the estimates
\[
\left\| {R\left( \gamma \right)} \right\| \le \frac{\left| {\gamma ^{ - 2}}
\right|}{\left| {Im\gamma ^{ - 2}} \right|} = \frac{\left| \gamma
\right|^2}{2\left| {\gamma }' \right|\left| {\gamma }'' \right|} =
\frac{1}{2}\left( {\frac{\left| {\gamma }' \right|}{\left| {\gamma }''
\right|} + \frac{\left| {\gamma }'' \right|}{\left| {\gamma }' \right|}}
\right) \le \frac{1}{2}\left( {1 + \frac{\left| \gamma \right|}{l}} \right)
\]
when $\left| {\gamma }'' \right| > \left| {\gamma }' \right|$ and $\left\|
{R\left( \gamma \right)} \right\| = 1$ for $\left| {\gamma }'' \right| \le
\left| {\gamma }' \right|$ where $\gamma = {\gamma }' + i{\gamma }''$.
Choosing the value $l > T_0 $ we obtain
\[
\left\| {\gamma ^{ - 1}T\left( \gamma \right)R\left( \gamma
\right)} \right\| \le \frac{1}{2}\left( {1 + \frac{\left| \gamma
\right|}{l}} \right)\frac{T_0 }{\left| \gamma \right|} < 1.
\]
Hence, there exists a bounded operator
\[
F^{ - 1}\left( \gamma \right) = R\left( \gamma \right)\left( {I + \gamma ^{
- 1}T\left( \gamma \right)R\left( \gamma \right)} \right)^{ - 1},
\]
which yields the existence of bounded operators $\tilde {L}^{ -
1}\left( \gamma \right)$ and $L^{ - 1}\left( \gamma \right)$ for
$\gamma \in \Pi _l = \left\{ {\gamma :\left| {Re\gamma } \right| <
l} \right\}$ outside the strip $\Pi _l $.


{\sc{Corollary 3.}}
\textit{The resolvent set of pencil} $L\left( \gamma
\right)$\textit{ is not empty,}
\[
{\bf C}\backslash \Pi _l \subset {\cal R} \left( L \right).
\]

\textbf{Theorem 2}. \textit{
\label{th2} The spectrum of pencil $L\left( \gamma \right)$ is
symmetric with respect to the real and imaginary axes:
\[
\sigma \left( L \right) = \overline {\sigma \left( L \right)} = - \sigma
\left( L \right).
\]
If $\gamma _0 $ is an eigenvalue of pencil $L\left( \gamma
\right)$ corresponding to the eigenvector $ \left( {\Pi ,\Psi }
\right)^{\rm T}$ then $ - \gamma _0 , \bar {\gamma }_0 $, and $ -
\bar {\gamma }_0 $ are also eigenvalues of pencil $L\left( \gamma
\right)$ corresponding to the eigenvectors $ \left( { - \Pi ,\Psi
} \right)^{\rm T},$ $  \left( {\bar {\Pi },\bar {\Psi }}
\right)^{\rm T}$, and $ \left( { - \bar {\Pi },\bar {\Psi }}
\right)^{\rm T}$ with the same multiplicity. }


{\bf{Proof.}} The first assertion of Theorem \ref{th2} follows
from (\ref{eq24w}) and (\ref{eq25w}). Proof of the second
assertion is actually a simple verification of variational
relation (\ref{eq14w}). Note that associated vectors at $\bar
{\gamma }_0 $ can be obtained by taking complex conjugation of
associated vectors corresponding to $\gamma _0 $.

\textbf{Theorem 3}. \textit{
\label{th3}
Set $\delta = \left( {\varepsilon
_2 - \varepsilon _1 } \right) / 2,$
\[
I_0 = \left\{ {\gamma :Im\gamma = 0,\frac{\left( {\delta ^2 + 4\varepsilon
_1 } \right)^{1 / 2} - \left| \delta \right|}{2} \le \left| \gamma \right|
\le \frac{\left( {\delta ^2 + 4\varepsilon _2 } \right)^{1 / 2} + \left|
\delta \right|}{2}} \right\}.
\]
In the domain ${\bf C}\backslash I_0 $ the spectrum of pencil
$\sigma \left( L \right)$ is a set of isolated eigenvalues with
finite algebraic multiplicity. The points $\gamma _j = \pm \sqrt
{\varepsilon _i }$ ($i = 1,2$) are the degeneration values of
pencil $L\left( \gamma \right):\dim \ker L\left( {\gamma _j }
\right) = \infty .$ }

{\bf{Proof.}} Taking $\gamma = \gamma ' + i\gamma ''$ with $\gamma
'' \ne 0$ we have
\[
Im\left[ {\frac{1}{\gamma ''}\left( {\gamma A_1 - 2\delta S -
\frac{\varepsilon _1 \varepsilon _2 }{\gamma }A_2 } \right)}
\right] = A_1 + \frac{\varepsilon _1 \varepsilon _2 }{\left|
\gamma \right|^2}A_2 \ge I;
\]
hence, in line with \cite{76}, the operator
\[
L_0 \left( \gamma \right): = \gamma ^2A_1 - 2\gamma \delta S - \varepsilon
_1 \varepsilon _2 A_2
\]
has a bounded inverse and $L\left( \gamma \right)$ is a Fredholm
pencil  with $\mbox{ind }L\left( \gamma \right) = 0$.

Introduce the operator $A'_1 :\,H \to H$ defined by the form ($\xi$ was defined before formula (\ref{eq17w}))
\begin{equation}
\label{eq33w} a'_1 \left( {f,g} \right): = \int\limits_\Omega {\xi
\left( {\varepsilon \nabla f_1 \nabla \bar {g}_1 + \nabla f_2
\nabla \bar {g}_2 } \right)dx} = \left( {A'_1 f,g} \right), \quad
\forall g \in H,
\end{equation}
and the operator
\[
\tilde {A}'_1 = A_1^{ - 1 / 2} A'_1 A_1^{ - 1 / 2} .
\]
For these operators the following estimates hold
\begin{equation}
\label{eq34w}
 - \varepsilon _{\max } I \le A'_1 \le \varepsilon _{\max } I,
\quad
 - I \le \tilde {A}'_1 \le I.
\end{equation}
Set $p = \left( {\left( {\varepsilon _2 + \varepsilon _1 } \right)
/ 2} \right)^{1 / 2}$.   For real $\gamma \notin I_0 $ the
estimate $\left| {\gamma ^2 - p^2} \right|
> \left| \delta \right|\left( {1 + \left| \gamma \right|} \right)$
holds; consequently, the operator
$$
\tilde {L}_0 \left( \gamma
\right): = A_1^{ - 1 / 2} L_0 \left( \gamma \right)A_1^{ - 1 / 2}
= \left( {\gamma ^2 - p^2} \right)I - 2\gamma \delta \tilde {S} +
\delta \tilde {A}'_1
$$
has a bounded inverse as well as operator $L_0 \left( \gamma
\right)$ . $L\left( \gamma \right)$ is a Fredholm operator with
zero index. Here, we used estimates (\ref{eq31w}) and
(\ref{eq34w}).

The second assertion of the theorem follows from variational
relation (\ref{eq14w}) for $\gamma = \gamma _j $ and
\[
\Pi ,\Psi \in C_0^\infty \left( {\bar {\Omega }_0 } \right),
\quad
\int\limits_\Omega {\Psi dx} = 0,
\]
for $\bar {\Omega }_0 \subset \Omega _1 $ and $\bar {\Omega }_0
\subset \Omega _2 $.

From the physical viewpoint the real and pure imaginary points of
spectrum $\sigma \left( L \right)$ are of interest because they
correspond to propagating and decaying waves. It should be noted
however that 'complex' waves may exist \cite{58, 71} for $\gamma
_0 \in \sigma \left( L \right)$ and $\gamma '_0 \cdot \gamma ''_0
\ne 0$ ($\gamma _0 = \gamma '_0 + i\gamma ''_0 )$. In the general
case, strip $\Pi_l $  in Theorem \ref{th1} cannot be replaced by
the set
\[
\Pi _0 = \left\{ {\gamma :\left( {Re\gamma } \right) \cdot \left(
{Im\gamma } \right) = 0} \right\}.
\]
From Theorem \ref{th2} it follows that complex waves occur in
`fours'. Note also that if a waveguide has a homogeneous filling
($\varepsilon _1 = \varepsilon _2 )$ then there are no complex
waves.

The existence of spectral points does not follow from Theorem
\ref{th3} (except for the points $\gamma _j = \pm \sqrt
{\varepsilon _i } )$. The proof of existence of a countable set of
eigenvalues of $L\left( \gamma \right)$ with an accumulation point
an infinity will be given below. Note that at the points $\gamma
_j = \pm \sqrt {\varepsilon _i } $,  the reduction of the boundary
value problem on normal waves to the eigenvalue problem for the
pencil is not valid (see Section 2.1). Hence one may expect the
occurrence of eigenvectors of pencil $L\left( \gamma \right)$
corresponding to eigenvalues $\gamma _j $. Using the methods of
potential theory \cite{104} we can prove that $L\left( \gamma
\right)$ is a Fredholm pencil  for all other real points $\gamma
$. Note that there are no other degeneration points and
finite accumulation points.

Let us prove the existence of discrete spectrum of pencil $L\left(
\gamma \right)$. First we prove the following statement.

\textbf{Lemma 4}. \textit{
\label{lem4} If the vector-function $\varphi \left( \gamma \right)
= F^{ - 1}\left( \gamma \right)\left( {\gamma ^{ - 1}f_0 + f_1 }
\right)$, $f_0 ,\,f_1 \in H$, is holomorphic for $\left| \gamma
\right| \ge R_0$ with a certain $R_0 > 0$ then this
vector-function is uniformly bounded (with respect to the norm) on
this domain. }

Let $\Lambda $ be the angle
\[
\Lambda = \left\{ {\gamma :\left| {\arg \gamma - \frac{\pi }{2}} \right| <
\theta ,\left| {\arg \gamma - \frac{3\pi }{2}} \right| < \theta } \right\}.
\]
Then for all $\gamma \notin \Lambda $ we have the estimates (see
the proof of Theorem \ref{th1})
\[
\left\| {R\left( \gamma \right)} \right\| \le \frac{1}{2}\left(
{\frac{\left| {\gamma }' \right|}{\left| {\gamma }'' \right|} +
\frac{\left| {\gamma }'' \right|}{\left| {\gamma }' \right|}}
\right) \le \frac{1}{2}\left( {1 + \cot \theta } \right),
\]
under the condition $\left| {\gamma }'' \right| > \left| {\gamma }'
\right|$. We also have
$\left\| {R\left( \gamma \right)} \right\| = 1$
for $\left| {\gamma }' \right| \ge \left| {\gamma }'' \right|$.
Thus
\begin{equation}
\label{eq35w} \left\| {R\left( \gamma \right)} \right\| \le 1 +
\cot \theta , \quad \gamma \notin \Lambda .
\end{equation}

Assume that $\left| \gamma \right| > R_1 > T_0 \left( {1 + \cot
\theta } \right)$ (value $T_0$ was defined in the proof of Theorem 1), $R_1 > R_0$ and $\gamma \notin \Lambda $. Then the
inequalities
\begin{equation}
\label{eq36} \left\| {F^{ - 1}\left( \gamma \right)} \right\| \le
\left\| {R\left( \gamma \right)} \right\|\left( {1 - \frac{T_0
}{R_1 }\left\| {R\left( \gamma \right)} \right\|} \right)^{ - 1}
\le \frac{1 + \cot \theta }{1 - \frac{T_0 }{R_1 }\left( {1 +
\cot \theta } \right)}
\end{equation}
follow from (\ref{eq35w}). Moreover, if the vector-function
$\mathord{\buildrel{\lower3pt\hbox{$\scriptscriptstyle\frown$}}\over
{\varphi }} \left( \gamma \right): = \gamma \varphi \left( \gamma
\right) = F^{ - 1}\left( \gamma \right)\left( {f_0 + \gamma f_1 }
\right)$, $f_0 ,\, f_1 \in H$, is holomorphic for $\left| \gamma
\right| = r > R_1 $ then (according to Lemma 1.3 in \cite{127})
\[
\ln \left\|
{\mathord{\buildrel{\lower3pt\hbox{$\scriptscriptstyle\frown$}}\over
{\varphi }} \left( \gamma \right)} \right\| \le c_1 \ln r + c_2
\int\limits_0^{c_3 r^2} {\frac{n\left( {t,\tilde {K}} \right)}{t}dt} ,
\]
where $n\left( {t,\tilde {K}} \right)$ is the number of $s$-values
of operator $\tilde {K}$ on interval $\left( {t^{ - 1},\infty }
\right)$. Since $\lambda _n \left( K \right) = O\left( {n^{ - 1}}
\right)$, we have \cite{76} $\lambda _n \left( \tilde {K} \right) =
O\left( {n^{ - 1}} \right)$, $n \to \infty $, and $n\left(
{t,\tilde {K}} \right) = O\left( t \right)$, $t \to \infty $. The
following inequality holds
\begin{equation}
\label{eq37w} \ln \left\|
{\mathord{\buildrel{\lower3pt\hbox{$\scriptscriptstyle\frown$}}\over
{\varphi }} \left( \gamma \right)} \right\| \le c_4 \left| \gamma
\right|^2.
\end{equation}

Choose $\theta < \frac{\pi }{8}$ and $R_1 > T_0 \left( {1 + \cot
\theta } \right)$. Estimates (\ref{eq36}) and (\ref{eq37w}) allow
us to apply the Phragmen--Lindeloeff principle \cite{130}
according to which
the boundedness of functions $\left\| {\varphi
\left( \gamma \right)} \right\|$ on the sides of angle
$\Lambda$, $\left| \gamma \right| > R_1 $ and the holomorphic
property of $\varphi \left( \gamma \right)$ yield the boundedness
of $\varphi \left( \gamma \right)$ for all $\left| \gamma \right|
> R_1 $ (including the points inside  angle $\Lambda $); the
following inequality holds
\begin{equation}
\label{eq38} \left\| {\varphi \left( \gamma \right)} \right\| \le
\max \left( {\frac{\left( {1 + \cot\theta } \right)\left( {\left\|
{f_0 } \right\|R_1^{ - 1} + \left\| {f_1 } \right\|} \right)}{1 -
\frac{T_0 }{R_1 }\left( {1 + \cot\theta } \right)},\mathop {\max
}\limits_{\left| \gamma \right| = R_1 } \left\| {\varphi \left(
\gamma \right)} \right\|} \right).
\end{equation}
Thus vector-function $\varphi \left( \gamma \right)$ is uniformly
bounded with respect to the norm in domain $\left| \gamma \right|
> R_0$.

\textbf{Theorem 4.} \textit{
\label{th4} In domain ${\bf C}\backslash I_0 $ the spectrum of
pencil $L\left( \gamma \right)$ forms a countable set of isolated
eigenvalues of finite algebraic multiplicity with an accumulation
point at infinity.}

{\bf{Proof.}}  It is sufficient to prove that the spectrum of
$L\left( \gamma \right)$ (or $F\left( \gamma \right))$ is not
empty in domain $\left| \gamma \right| > R$ for any $R$ (see
Theorem \ref{th3}).

Assume that the statement of the theorem is wrong. Then the
vector-function $\varphi \left( \gamma \right) = F^{ - 1}\left(
\gamma \right)f$ is holomorphic in domain $\left| \gamma \right| >
R$ for any $f \in H$. Take $R > \sqrt {\varepsilon _1 +
\varepsilon _2 } $ and assume that operator-function $F\left(
\gamma \right)$ has a bounded inverse on circumference $\left|
\gamma \right| = R$. From estimate (\ref{eq38}) it follows that
$\left\| {\varphi \left( \gamma \right)} \right\| \le c\left\| f
\right\|,$
$\left| \gamma \right| \ge R, \quad \forall f \in H,$
and
$\left\| {F^{ - 1}\left( \gamma \right)} \right\| \le c$,
$\left| \gamma \right| \ge R$.
Let us integrate the equality
\[
\left( {\gamma ^2\tilde {K} + I} \right)^{ - 1}f - F^{ - 1}\left( \gamma
\right)f = \left( {\gamma ^2\tilde {K} + I} \right)^{ - 1}\frac{1}{\gamma
}T\left( \gamma \right)F^{ - 1}\left( \gamma \right)f
\]
on arbitrary contour $\Gamma _k $ which contains inside only one eigenvalue $\gamma
_k $ of operator-function $\gamma ^2\tilde {K} + I$ (for example, $\Gamma _k :=
\{\gamma : |\gamma - \gamma_k| = \delta_k \} $ for sufficiently small $\delta_k > 0$). In this case
the integral of $F^{ - 1}\left( \gamma \right)f$ is equal to zero
and integrals of other terms are equal to the residuals at point
$\gamma _k $. Apply the expansion \cite{106} for resolvent
$R\left( \gamma \right)$ in the vicinity of $\gamma _k $
\[
\left( {\gamma ^2\tilde {K} + I} \right)^{ - 1} = \frac{1}{2\gamma _k
}\frac{1}{\gamma - \gamma _k }P_k + S_k \left( \gamma \right)
\]
(operator-function $S_k \left( \gamma \right)$ is holomorphic in
the vicinity of $\gamma _k $ and $P_k $ is an eigenprojector
corresponding to eigenvalue $\gamma _k )$. The expansion yields
\[
P_k \left( {I - \gamma _k^{ - 1} T\left( {\gamma _k } \right)F^{ - 1}\left(
{\gamma _k } \right)} \right)f = 0.
\]
In line with estimates $\left\| {T\left( \gamma \right)F^{ -
1}\left( \gamma \right)} \right\| \le T_0 c$, $\left| \gamma
\right| \ge R$, we obtain that for sufficiently large $\gamma _k $
(such $\gamma _k $ can be always chosen since eigenvalues of a
compact operator $\tilde {K} > 0$ have an accumulation point at
zero) the operator
$I - \gamma _k^{ - 1} T\left( {\gamma _k } \right)F^{ - 1}\left(
{\gamma _k } \right)$
has a bounded inverse. Since $f$ is arbitrary we have $P_k \tilde
{f} = 0$ for all $\tilde {f} \in H$. However $P_k \ne 0$, so that the latter
is not possible. This contradiction proves the theorem.

\begin{center}
\begin{figure}[htbp]
\centerline{\includegraphics[width=3.06in,height=3.01in]{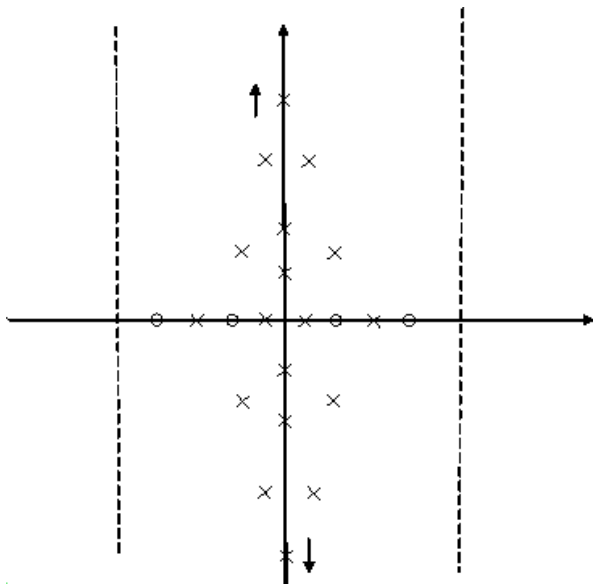}}
\caption{  \label{fig3} Spectrum of pencil $L\left( \gamma
\right)$ on the complex plane. {\rm o} and {\rm x} denote,
respectively, the degeneration points
and eigenvalues of pencil $L\left( \gamma \right)$
which are not equal to $\pm \sqrt {\varepsilon _i } $. }
\end{figure}
\end{center}

Figure \ref{fig3} shows the distribution of the spectrum of pencil
$L\left( \gamma \right)$ on the complex plane.



\section{Conclusion}

We have reduced the boundary eigenvalue problem for the Maxwell
equations describing normal waves in a broad class of
nonhomogeneously filled waveguides to an eigenvalue problem for an
operator pencil. We have proved fundamental properties of  the
spectrum of normal waves: the spectrum is nonempty and forms a
countable set of isolated points on the complex plane (cut along
two intervals on the real axis) without finite accumulation
points, is localized symmetrically in a strip, and contains not
more than a finite number of real points.
%
%
%

The results obtained in this work are of fundamental character for
the mathematical theory of wave propagation in guides and must be
used when particular types of nonhomogeneously filled waveguides
are considered in various applications.

\section{Acknowledgements}

This work is supported by the Visby Program of the Swedish
Institute.

\end{document}